# Comprehensive structural changes in nanoscale-deformed silicon modelled with an integrated atomic potential

Rafal Abram[a,†], Dariusz Chrobak[b,†], Jesper Byggmästar[c], Kai Nordlund[c], Roman Nowak[a,d,†,*]

[a] Nordic Hysitron Laboratory, School of Chemical Engineering, Aalto University, 00076 Aalto, Finland

[b] Institute of Materials Engineering, University of Silesia in Katowice, 41-500 Chorzów, 75 Pułku Piechoty 1A, Poland

[c] Department of Physics, University of Helsinki, P.O. Box 43, FI-00014, Finland

[d] Division of Materials Chemistry, Faculty of Engineering, Hokkaido University, Sapporo 060-8628, Japan

* Corresponding author: Roman Nowak, Email: roman.nowak@aalto.fi, Tel. +358 504602896
[†] These authors equally contributed to the present work.

## Abstract

In spite of remarkable developments in the field of advanced materials, silicon remains one of the foremost semiconductors of the day. Of enduring relevance to science and technology is silicon's nanomechanical behaviour including phase transformation, amorphization and dislocations generation, particularly in the context of molecular dynamics and materials research. So far, comprehensive modelling of the whole cycle of events in silicon during nanoscale deformation has not been possible, however, due to the limitations inherent in the existing interatomic potentials. This paper examines how well an unconventional combination of two well-known potentials - the Tersoff and Stillinger-Weber - can perform in simulating that complexity. Our model indicates that an irreversible deformation of silicon (Si-I) is set in motion by a transformation to a non-diamond structure (Si-nd), and followed by a subsequent transition to the Si-II and Si-XII' phases (Si-1→Si-nd→Si-II→Si-XII'). This leads to the generation of dislocations spreading outwards from the incubation zone. In effect, our simulations parallel each and every one of the structural changes detected experimentally in the deformed material. This includes both the sequence of phase transitions and dislocation activity, which - taken together - neither the Tersoff nor Stillinger-Weber, or indeed any other available Si interatomic potential, is able to achieve in its own right. We have sought to additionally validate our method of merging atomic potentials by applying it to germanium, and found it can equally well predict germanium's transformation from a liquid to amorphous state.

# 1  Introduction

*1.1 Mechanical behaviour of silicon*

Silicon is a material with a superior blend of physical properties and well-established means of fabrication, finding application in a vast range of fields [1] from electronics and photonics to micro-and nano-electromechanical systems, to name but a few. Early research into silicon discovered its exceptional mechanical properties, and has been widely published [2]. In particular, pioneering hardness experiments by Giardini [3], Gerk and Tabor [4], Petersen [5] as well as Gilman [6,7] revealed silicon's unorthodox mechanical behaviour. The mechanical response of silicon differed from that of other materials by having its plastic deformation rely on pressure-induced transformation from a diamond (Si-I) to b-tin (Si-II) structure [4,6,7], accompanied by dislocation mechanisms characteristic of crystalline solids (*e.g.*, Ref. 6). This has led to a close investigation of stress-induced metallization and amorphization of an indented Si-crystal [8].

With the advent of depth-sensing nanoindentation technology [9-11], hardness experiments on silicon reached a new level. Early results by Pethica *et al.* [11] and Pharr *et al.* [12] and a rapid development of the nanoindentation method led to the recognition of the complexity of silicon nanodeformation, by contrast with other ceramics or metallic crystals [13-15]. Research into the role of phase transformations in nanodeformed silicon has continued (*e.g.*, Refs. 15-19), and seems far from having reached its saturation point (see *e.g.*, Refs. 18-26), being increasingly supported by atomistic modelling [27-31].

*1.2 Atomistic simulations*

Modern materials science owes much of its success to computer simulations [32], which examine and predict a large array of phenomena (*e.g.*, [24-30]). The molecular dynamics (MD) simulation method accounts for the interatomic forces in terms of the gradient of the potential energy function in conjunction with a set of adjustable parameters to predict the specific conduct of a modelled system. Inevitably, whether or not the outcome of simulation reflects a given set of experimental data depends on the adopted interatomic interaction model [32,33]. Materials physics has no shortage of "ready-made" potentials at its disposal, however, their limited transferability means that one process is usually modelled at the expense of another, thus precluding the interplay of key phenomena. Indeed, any model of atomic interactions involves a trade-off between accuracy, adaptability and computational resources [32]. Thus, a careful consideration of the available interatomic potentials focusing transferability precedes any MD simulations [34,35].

Often, after deciding that none of the applied potentials mirrors a system's behaviour of interest, developing a suitable one 'from scratch' becomes a necessity. It is a complex and time-consuming trial and error process. Here we examine a simple and promising alternative: it is to merge two already existing - yet operationally diverse – potentials developed for Si crystal, into a single one. This new, integrated potential (IP) - by combining the unique characteristics of its components - makes it possible to simulate some crucial, coexisting features of a system under examination. Using the IP-approach, we have been able to deal with the intricate process of nanoindentation deformation of silicon in a more satisfactory way than appears possible with the currently available potentials [32,33].

*1.3 Structure evolution in nanodeformed Si*

As already mentioned, silicon is probably the most extensively studied semiconductor in modern technology. Its applications include surface acoustic wave photonic devices [36], light sources [37] or atomic scale instruments and memories [38]. One can therefore be forgiven for believing that every property of this important material has been thoroughly investigated [39]. But the fact of the matter is that such phenomena as nanoscale deformation of silicon crystal, and its incipient plasticity, in particular, are far from being fully understood. Numerous nanoindentation experiments, combined with microscopic observations, demonstrate that, under increasing straining, the initial Si-I structure undergoes a series of transformations to high-pressure Si-II phase. The latter converts during the unloading to the Si-III/Si-XII combination and amorphous Si phases [22,40-47]. This, however, is not the complete picture as it disregards the presence of dislocations, slip bands and cracks in the vicinity of the acting indenter tip [17,20,22,45,48]. A microscopic examination by Wong *et al.* [22,43-46] of the Si structure developed under a spherical indenter led them to the conclusion that phase transformation and defects generation were "competing deformation mechanisms, with one or the other process initiating plastic deformation under particular loading conditions". Similarly, nanoindentation-induced elastic-plastic transition in GaAs crystal is initiated by a phase transformation [49-51], while nanoscale plasticity in GaN is exclusively governed by dislocations [52,53].

What we are dealing with here are two distinct mechanisms of silicon's incipient plasticity: in the first, phase transformation assumes the dominant role followed by dislocation activity, while in the second, plasticity starts with the nucleation of defects. Regardless of which scenario unfolds in practice, none of the currently available potentials meets the requirement of accurately modelling a complex nanoindentation process, since each one of them suffers either from limited transferability [54] or high computational-time expense. Such is the case with the interatomic potentials developed with the aid of machine learning [55,56] which are capable of a high degree of accuracy, even on a par with *ab initio* methods. However, they frequently exhibit such computational complexity as to render them impracticable for the purpose of nanoindentation, which requires a simultaneous consideration of multimillion atomic systems and the application of long-time scales [33]. As a result, modelling silicon mechanical behaviour continues to rely heavily on either the Stillinger-Weber (SW) [57] or Tersoff-type [58-65] potentials. The former has been designed with the Si-I and amorphous phase in mind, proving particularly efficient at capturing dislocation-based processes. By contrast, Tersoff-type potentials reflect phase transformations that occur in the high-pressure phases (diamond cubic and metallic) of silicon. It follows that neither model is versatile enough to accurately reflect the interplay between both phenomena, thus precluding proper investigation of Si crystal behaviour under localized stress.

An early critical appraisal of the interatomic potentials for Si has been published by Balamane *et al.* [66], while tests on the Si lattice under large shear strain have been performed by Godet *et al.* [67]. Furthermore, there have been numerous efforts to modify the original Tersoff interatomic potential in order to increase its accuracy and transferability, *e.g.* [62]. Kumagai *et al.* [64] revised the angular-dependent term of the Tersoff formula, considerably improving the description of silicon's elastic properties as well as melting temperature. In consequence, the Kumagai's potential has recently been employed to model nanoindentation-induced deformation of silicon [68,69]. The most significant advance in the simulation of silicon brittle behaviour has been achieved by Pastewka *et al.* [36], whose screened version of the Erhard-Albe potential (T3s) [63] manages to capture both phase transformations and dislocation activity in silicon [70-72]. Even the most advanced, the screened Tersoff T3s potential results in an overestimation of contact pressure at the onset of phase transformation, as well insufficiently developed dislocation structure [71]. The overall conclusion is that crystalline phase transformations in silicon are usually modelled at the expense of dislocation movements, or vice versa [47,73] and a more adaptable interatomic potential is required.

To recapitulate, the integrated interatomic potential (IP) of our design, obtained by simply merging the Tersoff and Stillinger-Weber models, appears well-suited to MD-modelling of the structural changes in nanodeformed Si-crystal, proving it inherits the essential features of the component potentials. Our results provide a consistent account of the sequence of phase transformations and dislocation activity, both of which play a crucial role in nanoindentation-induced plasticity of silicon. Finally, our findings concerning the structure evolution in a stressed nano-volume of silicon — by virtue of being firmly rooted in MD simulations — are in line with the recent research into silicon, whether dealing with the structure and stability of the interstitial defects [74], the amorphization by mechanical deformation [75,76], shuffle dislocations motion in Si-crystals [25,77] or low-temperature undissociated-dislocations-mediated plasticity [31].

## 2. Computational

### 2.1 The integrated potential for Si

Our [47,73] own experience and that of other authors (*e.g*., Ref. 78) shows that satisfactory modelling of Si behavior exclusively with the Tersoff (T2) potential is limited to phase-transformation governed phenomena. A similar problem arises with the SW potential, whose competence is limited to deformation processes steered by dislocation-activity [73]. Faced with this dilemma, we have attempted to merge the Stillinger-Weber and Tersoff potentials - two of the most acclaimed interatomic potentials for silicon - into a single model in order to obtain one of enhanced transferability.

The solution we offer consists of the integrated potential (IP), based on a weighted sum of the SW and T2 (Tersoff) potentials, such that the potential energy $E^{IP} = w_{T2}E^{T2} + w_{SW}E^{SW}$ ($w_{T2} + w_{SW} = 1$), where the $E^{T2}$ stands for potential energy term (equation T2 – Ref. 59), incorporates the effect of the local atomic environment, while the $E^{SW}$ (equation SW – Ref. 57) accounts for both two- and three-body interaction terms (Supplementary Note 1). Careful testing of variously weighted versions of our integrated potential resulted in a unique IP-combination with the weight values $w_{T2} = 0.71875$ and $w_{SW} = 0.28125$, which we regard as close to optimal for silicon under localized stress.

Although the physical basis of the T2 and SW potentials differs, each one can be viewed as an angle-dependent, embedded-atom potential [79]. The purpose of complementing the T2 model with a smaller share of the SW potential was to "stiffen" the response of the former. It needs reminding that the SW potential contains a penalty function strongly suppressing deviation from the bond angle of 109.47°, characteristics of diamond structure.

## 2.2 Initial testing of integrated IP potential for Si

Initial testing of IP performance in simulating silicon was carried out using the T2 and SW potentials, alongside the T3s (screened) models, for comparison. It concerned potential energy $E(r)$ and its derivative ($dE/dr$) calculated as the function of the interatomic distance $r$ in the Si-Si dimmer, followed by an evaluation of the elastic constants, cohesion energy and the energy of an unstable glide-set stacking fault. All the computations were accomplished with the LAMMPS simulation package [80], while the calculations with the screened T3s potential required the employment of a code from the library of interatomic potentials ATOMISTICA [81]. Our simulations involved a super-cell (4096 atoms) composed of 8×8×8 Si unit cells with the lattice parameter $a$=5.431 Å, and a structure optimized using the conjugate gradient method. The cohesion energy was defined as a ratio of the total potential energy and the number of atoms, while the elastic constants were estimated by applying small volumetric, tetragonal and rhombohedral deformations to the Si supercell. Finally, the stacking fault energy calculations followed those proposed by Branicio *et al.* [82].

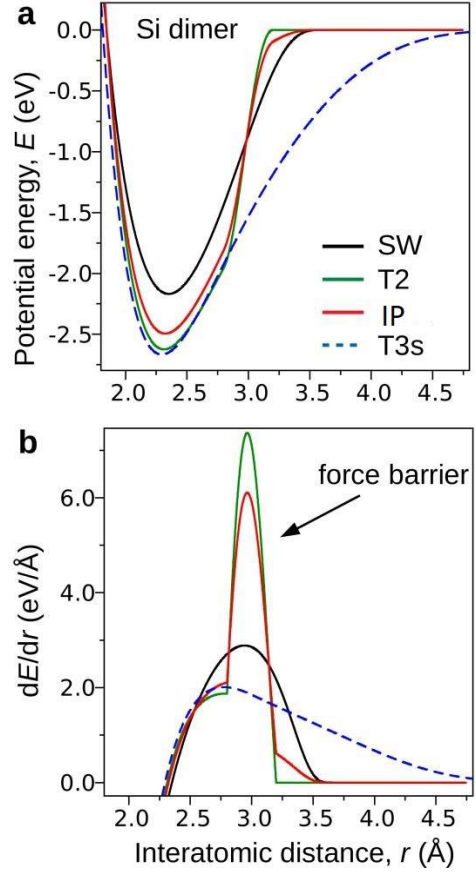

**Fig. 1.** The potential energy $E(r)$ of a two-body interaction between the atoms of the Si-Si dimmer determined with the IP (red), SW (black), T2 (green) and T3s (blue) potentials (a) and its derivative $\mathrm{d}E/\mathrm{d}r$ functions (b). The displayed results concern the Si-Si dimmer with the interatomic distance $r$ and simulation conditions defined by the penetration depth $h$=18 Å and contact pressure $p_c$=7.7 GPa.

Our initial results prove that the IP potential inherits the characteristics of both the T2 and SW potentials, as well as showing considerable promise for modelling the structural changes in Si under complex stress Fig. 1(a). It is worth noting that the energy curve obtained with the IP potential (red) is located in-between those obtained with the T2 (green) and SW models (black), its exact position depending on the combination of the $w_{T2}$ and $w_{SW}$ parameters. On the other hand, we found that the IP simulation delivered a marginally lower Si-Si dimmer's binding energy than the T3s (Fig. 1a). It was also characterised by a smaller range, leading to a shorter simulation time. Importantly, MD calculations with the IP potential involved only one coordination shell, while those with T3s involved two. The decrease in the force-barrier of the Si-Si dimmer (Fig. 1b) suggests the IP has preserved the Tersoff potential attributes in addition to its ability to describe dislocation activity. This contrasts with earlier suggestions [65] that the force-barrier prevents the modelling of irreversible deformation, including fracture and plasticity.

A further comparative examination of the IP, T2, SW and T3s interatomic potentials' performance, when applied to silicon, includes the elastic constants ($c_{11}$, $c_{12}$, $c_{44}$), cohesive energy ($E_c$), and stacking fault glide energy ($u_{<112>}$), whose values have been listed in Table 1.

**Table 1** Cohesive energy $E_c$ [eV], elastic constants $c_{ij}$ [GPa] as well as relaxed stacking fault energy ($u$-unstable, J/m$^2$) calculated with the T2, SW, IP and T3s potentials. The lattice constant of the Si-I phase is 5.431 Å, regardless of the potential selection. The weighted sum $w_1X_1 + w_2X_2$ of the considered properties of silicon was also considered.

| | T2 $X_1$ | SW $X_2$ | $w_1X_1 + w_2X_2$ | IP | T3s | Published data |
|---|---|---|---|---|---|---|
| $E_c$ | -4.63 | -4.34 | -4.55 | -4.55 | -4.63 | -4.62 [65] |
| $c_{11}$ | 122 | 151 | 130 | 130 | 143 | 166 [65] |
| $c_{12}$ | 86 | 76 | 83 | 83 | 75 | 64 [65] |
| $c_{44}$ | 10 | 57 | 23 | 24 | 69 | 80 [65] |
| $u_{<112>}$ | 1.84 | 3.20 | 2.22 | 2.22 | 2.20 | 191 [67] 256 [83] |

It seems quite remarkable that the results obtained for IP are close to the weighted sum value of those calculated with the T2 and SW potentials, a minor exception being the $c_{44}$ elastic constant (Table 1). This serves as a further indication of IP validity. Since the shear modulus value of $c_{44}$ = 24 GPa obtained with IP is higher than the one with T2 ($c_{44}$ = 10 GPa), it can be applied to realistic modelling of both phase transformations as well as dislocation activity during the nanoindentation of Si crystal.

In summary, the initial examination of the IP model alongside other potentials found that our approach — although deceptively straightforward — shows considerable promise in capturing the structural changes under complex stress. At the same time, it has to be admitted that, as far as the elastic properties of Si are concerned, the screened Tersoff T3s [65] potential tends to be slightly more accurate in its reflection than our IP model.

*2.3 MD-simulation of indentation in Si*

The initial testing of the integrated IP potential was followed by 3-way MD-simulations of the spherical nanoindentation of the Si crystal in order to find out which of the employed potentials - IP, T2 or T3s - best reflected the structural changes known to occur under experimental conditions [20,22,40-46,48,73]. Excluded from the simulations was the SW potential, which does not capture pressure-induced transitions to crystalline phases of silicon [73].

Our MD simulations were performed with the LAMMPS simulation code for a silicon cluster (30.7 nm × 30.7 nm × 27.2 nm, 1328142 atoms) whose (001) surface was deformed under contact with a rigid, spherical diamond tip (radius of $R$ = 16.3 nm). The coordination axes (X, Y, Z) were aligned along the [110], [-110] and [001] Si crystal directions respectively. Furthermore, the standard velocity-Verlet time integration scheme, with a time-step of 2 fs, was used throughout, while the Nose-Hoover thermostat was employed to control the system. Prior to running nanoindentation simulations, the system was relaxed to a thermal equilibrium at a target temperature of 300 K. Interactions between the diamond indenter tip and silicon crystal were modelled using the repulsive term of the Buckingham potential (the cut-off radius of 4 Å). MD-simulated nanoindentation of the Si crystal was accomplished by a sequence of tip displacements, with an increment of 0.5 Å every 15 ps, thus securing a quasi-static crystal deformation.

The defected structure generated in the deformed Si crystal was examined in detail using tools (modifiers) incorporated into the OVITO software [84,85]. In order to obtain a clear presentation of the structural changes in the Si crystal caused by nanoindentation, atoms belonging to the diamond structure have been disregarded. The applied procedure (*Identify diamond structure* modifier) leaves only atoms whose arrangement corresponds to the non-diamond (high-pressure) phases of silicon. Moreover, the atoms have been attributed by the values of the shear strain (*Atomic strain* modifier). The identification of dislocations and their Burgers vectors was accomplished using the dislocation extraction algorithm (*DXA* modifier) [86].

Our findings were verified by means of a visualization of the structural changes in nanoindented silicon, and by employing the bond-angle distribution (BADF) and radial distribution functions (RDF) to identify high-pressure silicon phases. The BADF was calculated by determining the bond angles in spheres with a radius 3 Å, centred on each Si atom. The range of bonding angles, 0°-180°, was divided into 200 sections. Then the number of bond angles $N_i$ ($i = 1, ... ,200$) belonging to the specific $i$-th interval was calculated. As the groups of particles selected for a structural analysis did not contain an equal number of atoms, we used normalized distribution $N_i/N$, where $N$ stands for the number of atoms in a group under analysis. A similar procedure was used in the case of the RDF, with the interatomic distances investigated within the spherical volume of 5 Å radius.

## 3. Results

*3.1 MD-simulations of indentation in Si with the Tersoff T2 potential*

Our MD-simulations with the Tersoff T2 potential revealed an elastic/plastic transition in the original Si-I structure at the contact pressure of $p_c$=4.2 GPa and indentation depth of $h$=13.5 Å (Fig. S2a). The observed incipient plasticity (Supplementary Note 2, Fig. S1) is a phase transformation–governed phenomenon, similar to the one we have found in a nanodeformed GaAs crystal [50,51], and quite unlike the dislocation-governed incipient plasticity of the InP crystal [87]. Interestingly, the Si-I→Si-II' phase transition begins and continues directly under the loaded spherical indenter at a certain distance from the contact surface. The seed of the Si-II' phase shows up exactly in the area of enhanced shear strain (Fig. 2 and Fig. S2). As indentation proceeds, the nucleus continues to grow until well-defined columns of the Si-II' phase are formed, as illustrated in a 20 Å thin section across the region deformed under the contact pressure of $p_c$=7.4 GPa and the contact depth of $h$=39 Å (Fig. 2a).

The Si–II' phase appears under the acting indenter in the form of vertical zones of different orientation, shown in Fig. 2a by successive numbers 1~6 (regions 1, 3 and 5 are rotated by approximately 10.5° in relation to sectors 2, 4 and 6). What we also find is that the tetragonal unit cell of the high-pressure Si-II' phase (Fig. 2b) is a distorted (c/a≈1) version of the well-known Si-II (β-Sn, c/a=0.55) structure [88-90].

The structural difference between the Si-II' and Si-I phases is made explicit by the bond angle (BADF) and radial (RDF) distribution functions (Fig. 2c). The BADF analysis applied to Si-II' in different vertical zones (Fig. 2a) as well as to the initial Si-I phase reveals the distinct Si-I peak at 109º, which promptly splits into two (100º and 145º) as indentation progresses (Fig. 2c). Furthermore, the RDF function confirms that the structure transformation did indeed take place (Fig. 2c). The sequence of characteristic peaks disclosed at the interatomic distance of 2.35, 3.84, and 4.50 Å, involves the Si-I structure, while the spectrum 2.35, 3.18, 3.58, 4.23, and 4.50 Å defines the Si-II' phase.

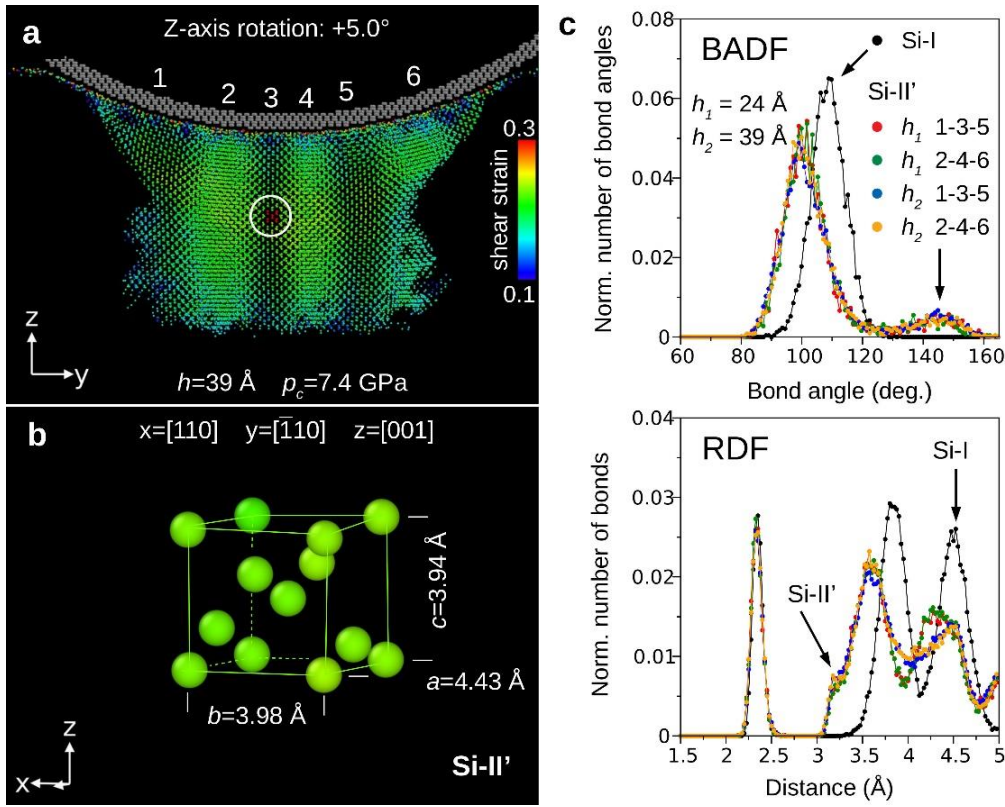

**Fig. 2.** Modelling with the T2 potential. (a) The Si-II' structure detected in a 20 Å thin section along the axis of indenter symmetry (z), across the transformed zone. Columns of the Si-II' structure, labelled 1 ~ 6, differ in orientation around their vertical axes by approx. 10.5°, while the colour of the Si atoms signals the atomic shear strain level. (b) The tetragonal unit cell detected in the circled area (a) points to a highly compressed Si-II structure. (c) BADF and RDF functions disclose characteristics of Si-I and Si-II' phase, determined at $h_1$=24 Å and $h_2$=39 Å indentation depths, respectively (c).

In summary, MD-simulations with the T2 potential succeed in modelling the transformation of Si-I to Si-II' phase, which corresponds to the experimentally determined [90] Si-II. However, the T2 fails to capture any dislocation activity in silicon across the whole range of depth that we have considered.

*3.2. MD-simulations of Si nanoindentation with the T3s potential.*

The structure evolution of an Si crystal under a spherical diamond tip, begins with the nucleation of a tiny seed of a high-pressure phase at the indentation depth of $h$=14.5 Å and contact pressure $p_c$=12.6 GPa (Fig. S1). This initial transformation takes place at a distance of approx. 25 Å from the contact area, pointing to the dominant role of shear stress in the formation process. However, once the indentation depth reaches $h$=20 Å, under the contact pressure of 15.9 GPa, the volume of a new high-pressure phase (BCT-5') is sufficient to allow us its reliable identification (Fig. 3a and Fig. 3b), which turns out to be a distorted version of the BCT-5 structure [88,91].

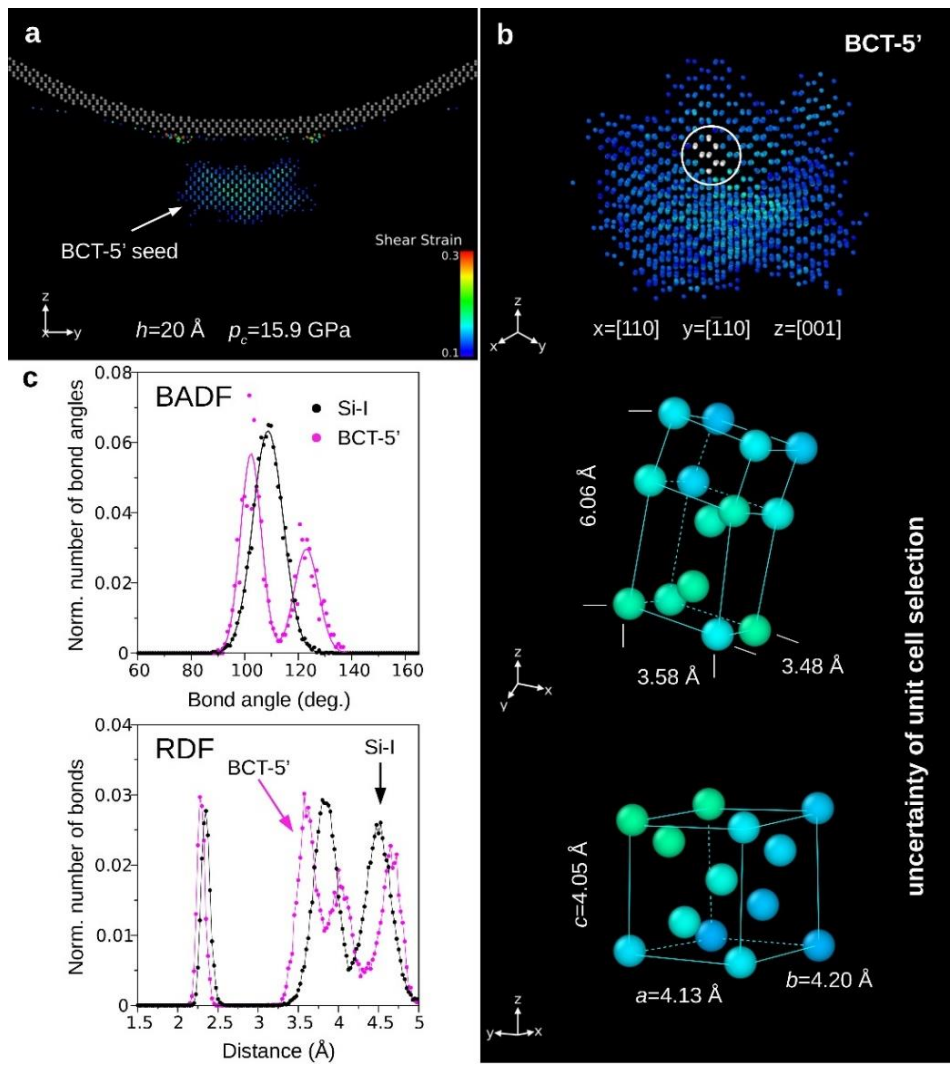

**Fig. 3.** Modelling with the T3s potential (atom colour specifies shear strain level). (a) High-pressure phase BCT-5' positioned approx. 30Å below the contact area. (b) Si structure rotated to demonstrate atom arrangement of BCT-5' area marked by open circle (b). The presented unit cell shows deviation from BCT-5 structure [73]. BCT-5' lattice is commensurate with compressed tetragonal Si-I phase. (c) BADF and RDF functions at the $h$=20 Å indentation depth, reveal structural differences between BCT-5' and Si-I phases.

As a matter of fact, the lattice of the BCT-5' phase structure is by no means unique, as it might also be described as a highly compressed tetragonal unit cell of the initial Si-I phase (Fig. 3b). A highly-compressed Si-I tetragonal lattice has the ratio approaching unity, in contrast to the ideal, strain-free crystal for which c/a =$\sqrt{2}$. A good illustration of our findings is provided by the BADF and RDF functions whose peak sequences, characteristic of the BCT-5' phase, equal 102°, 123° and 3.81, 4.0, 4.68 Å, respectively (Fig. 3c). The results are in contrast to those obtained using the T2 potential (Fig. 2c).

Further consideration of our MD-simulated nanoindentation has brought us to the conclusion that the growth of the BCT-5' phase is significantly affected by an increase of shear strain within its volume (Fig. 4). The partial disappearance of the largely unstable BCT-5' phase ($p_c$=17.8 GPa, Fig. 4a) results in the formation of shear bands along the {111} planes followed by the development of Si-II and BCT-5 phases (Fig. 4a), whose characteristics broadly agree with earlier experiments and calculations [88-91]. The main difference between the unit cells of the BCT-5' and BCT-5 (Fig. 4b) consists in the former possessing a larger base than the latter, with a characteristic distortion in its upper part. Contrary to common belief, the Si-II phase shows up within the shear bands oriented along the {111} planes (Figs. S3) rather than in the material adjacent to the contact area [22,44].

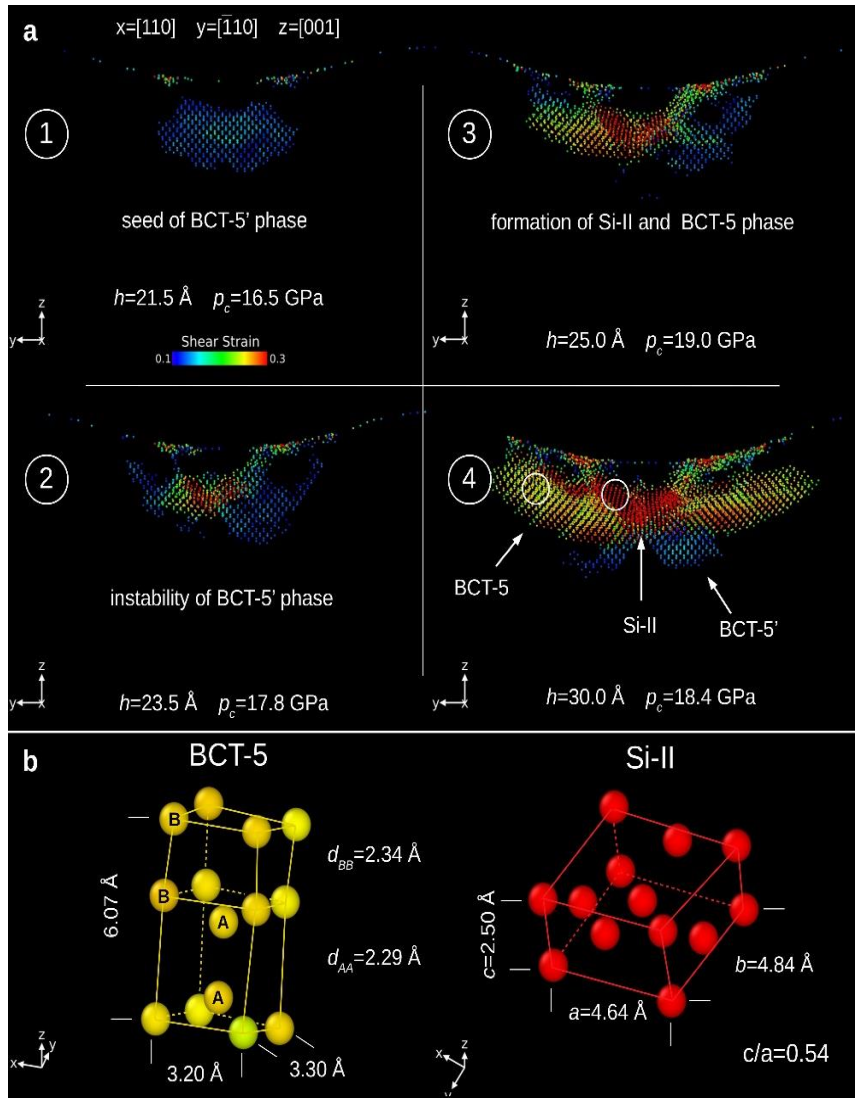

**Fig. 4** MD-simulations with the T3s potential: Evolution of high-pressure BCT-5' and Si-II phases. (a) Selected indentation stages 1~4, showing shear strain development in the BCT-5' phase (snapshot 1). The particular atom colour signals either shear strain level, or the Si-II (red) and BCT-5 (yellow) phases. The {111} shear bands are formed in the late stages of indentation (snapshot 4).   (b) BCT-5 and Si-II phases from snapshot 4 shown in focus as unit cells.

Our visualization of the evolution of a strained atomic Si structure accomplished with the DXA-dislocation extraction algorithm [86] has enabled us to examine closely dislocation activity, which always accompanies nanoindentation in Si and has been reported widely on the basis of TEM cross-section observations [17,20,22,43-46]. It is an advantage of the T3s potential to be able capture dislocation processes in tandem with phase transformations [70-72]. However, the DXA analysis has revealed but a few dislocations on the {111} planes (the Burgers vector $\left|\vec{b}\right| = \frac{1}{2}\langle110\rangle$) which develop parallel to the (001) surface (Fig. 5). Furthermore, we have found no 'downward defects', *i.e.,* moving in the direction opposite to the surface, despite numerous TEM reports of their existence [22,44,48].

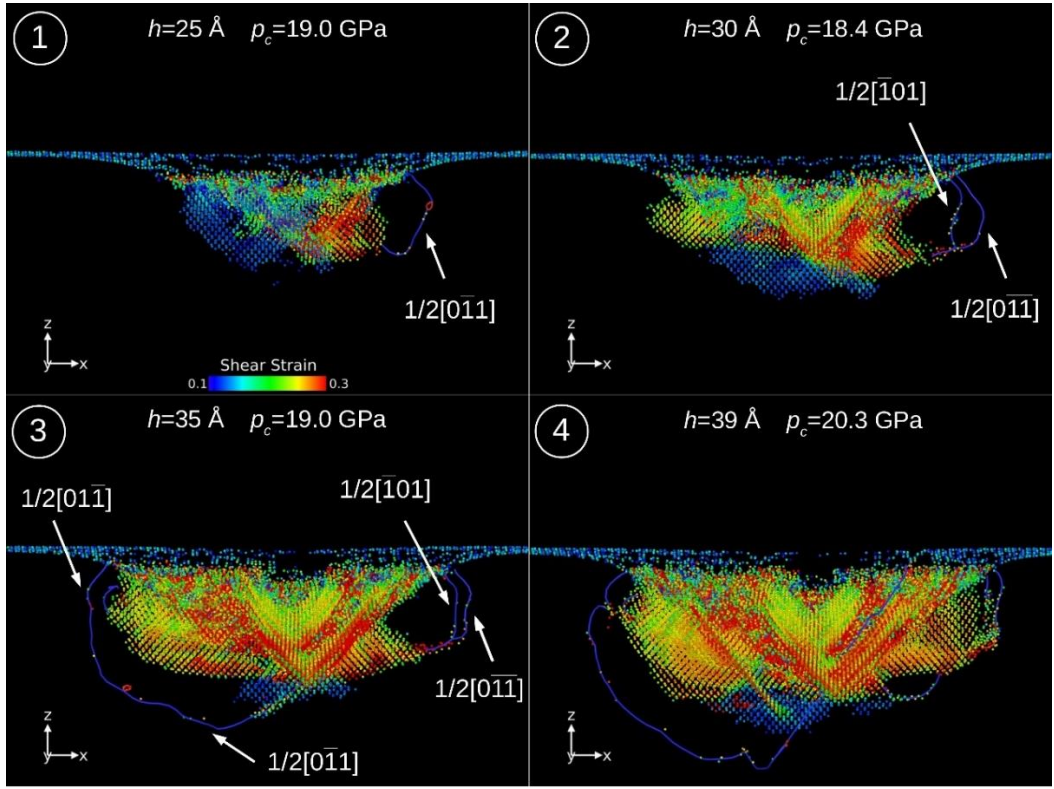

**Fig. 5** MD-simulations with T3s potential: The evolution of dislocation structure in the nano-deformed Si. The sequence of structure-snapshots (from 1~4) taken for respective indentation depths of *h*=25, 30, 35 and 39 Å provide the DXA based visualization of dislocation structure and its development. The defects (the Burgers vector defined in the figure) are generated in the {111} planes, outside of the regions occupied by high-pressure Si phases.

All in all, while the results of our MD-simulations with the T3s potential agree with those obtained with the EAs and T3 potential [71,88,89], we have found that using the T3s approach in Si nano-deformation is not without its drawbacks. These concern primarily the contact pressure necessary to initiate the Si-I→Si-II transformation (~18 GPa), which is much overestimated compared to the ~11 GPa level measured in anvil cell experiments [92], or ~9.9 GPa derived from nanoindentation tests [48]. In addition, the Raman spectroscopy investigations undermine the possibility of the BCT-5 phase nucleation taking place at all during Si nanoindentation. Indeed, recent research [41,42] points to the existence of a distorted diamond cubic structure (*dc-2*) instead of the BCT-5, in contrast to the results of simulations with the T3s potential (Fig. 4).

To conclude, a visualization of nano-deformed Si crystal shows that nanoindentation induced changes modelled with the T3s screened potential are initiated by the Si-I→BCT-5' transformation that precedes the formation of the Si-II and BCT-5 phases. The predicted dislocation structure is largely underdeveloped compared to microscopic observations [48].

### 3.3 MD-simulations with the integrated SW-T2 (IP) potential

The irreversible deformation of Si crystals begins with the nucleation of a tiny BCT-5' phase seed (at the depth $h$=16 Å and pressure $p_c$=7.9 GPa - Fig. 6a, snapshot 1), analogously to the forecast made with the T3s potential (Fig. 4a - snapshot 1). In the same way, simulations with the IP potential, support the prediction that the nucleation of the BCT-5' phase occurs precisely at the zone of an increased shear strain. As the indentation process unfolds, the detected high-stress structure grows, reaching a stage ($h$=20.5 Å, $p_c$=8.2 GPa) when it undergoes the BCT-5→Si-II' transition (Fig. 6a - snapshot 2). The resulting Si-II' expands further, forming a specific, compact seed of considerable volume adjacent to the contact area, in direct confirmation of the TEM observations by Wong *et al.* [22,43,44]. This very form of the new phase, with its cohesion and particular location, points towards effectiveness of the IP potential (compare Fig. 4a, Fig. 6a, Fig. S3 and Fig. S4). Furthermore, as straining increases, part of the material with the Si-I structure, transforms into another crystalline structure, namely Si-III' (Fig. 6a, snapshots 3-4). This process takes place directly beneath, and outside, of the formed Si-II'.

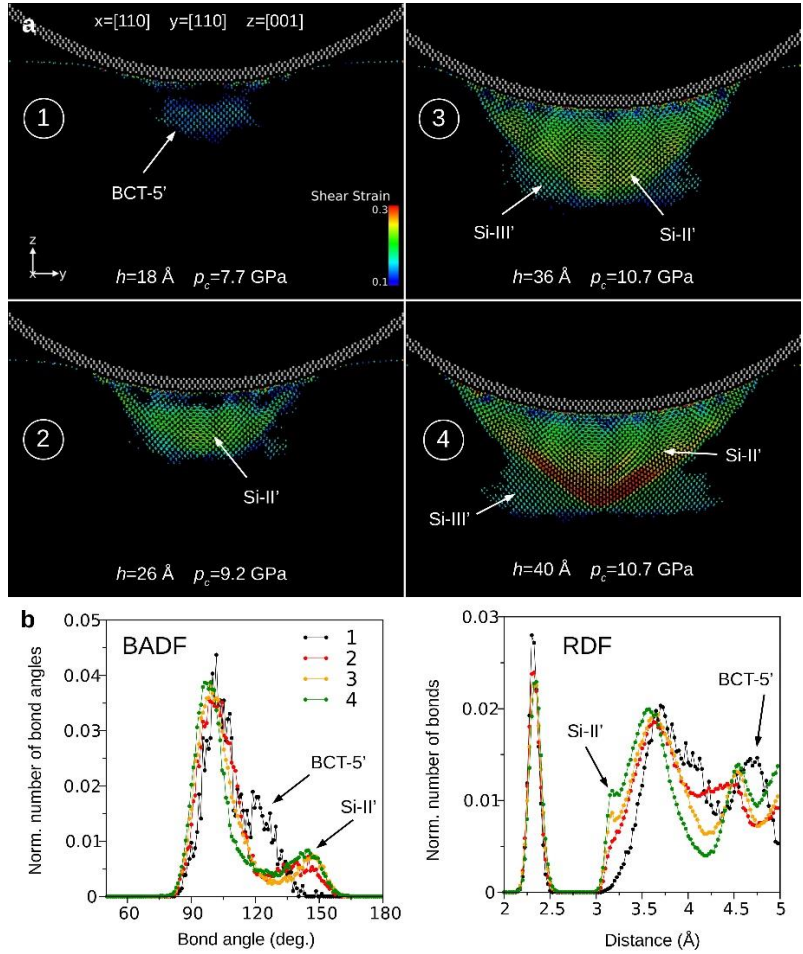

**Fig. 6.** Modelling with the integrated IP potential: The sequence of indentation-induced phase transformations in Si. (a) Structure evolution during consecutive indentation stages: $h$=18, 26, 36 and 40Å. The atoms of the original Si-I phase are not displayed. The 20 Å thin section across the BCT-5' phase (snapshot 1) and the Si-II' phase (snapshots 2 and 3) illustrate the structure evolution and formation of the Si-III' phase (snapshots 3 and 4) either directly under or in the vicinity of the Si-II' volume. (b) BADF and RDF functions for conditions in snapshots 1~4 excluding the Si-III structure.

The BADF and RDF analysis (Fig. 6b) of the atom arrangements, displayed in the consecutive snapshots (Fig. 6a), corroborates the sequence of crystal structure transformations that we have presented. Moreover, there is a perceivable resemblance between the BADF and RDF atomic distributions (excluding the Si-III' phase) derived with our IP potential for the Si-II' structure (Fig. 6b) and those obtained by means of the T2 potential (Fig. 2c). The same applies to the BCT-5' phase, which the IP computer experiments have revealed in a similar way as has the T3s potential (compare Fig. 6b and Fig. 3c). This allows us to conclude that the integrated IP approach - by combining the strengths of both the T2 or SW potentials - is more universal than any of the three - T2, SW and T3s.

In order to provide a detailed insight into the results obtained with the IP potential, we have scrutinized the atomic arrangements of the BCT-5' and Si-II' phases (Fig. 7). This shows the tetragonal parameter of the Si-II' phase achieving its lowest value of c/a≈0.7. Moreover, the Si-II' structure in the proximity of the contact surface displays a higher value of c/a ratio (approx. ~0.88 refer to Fig. S5).

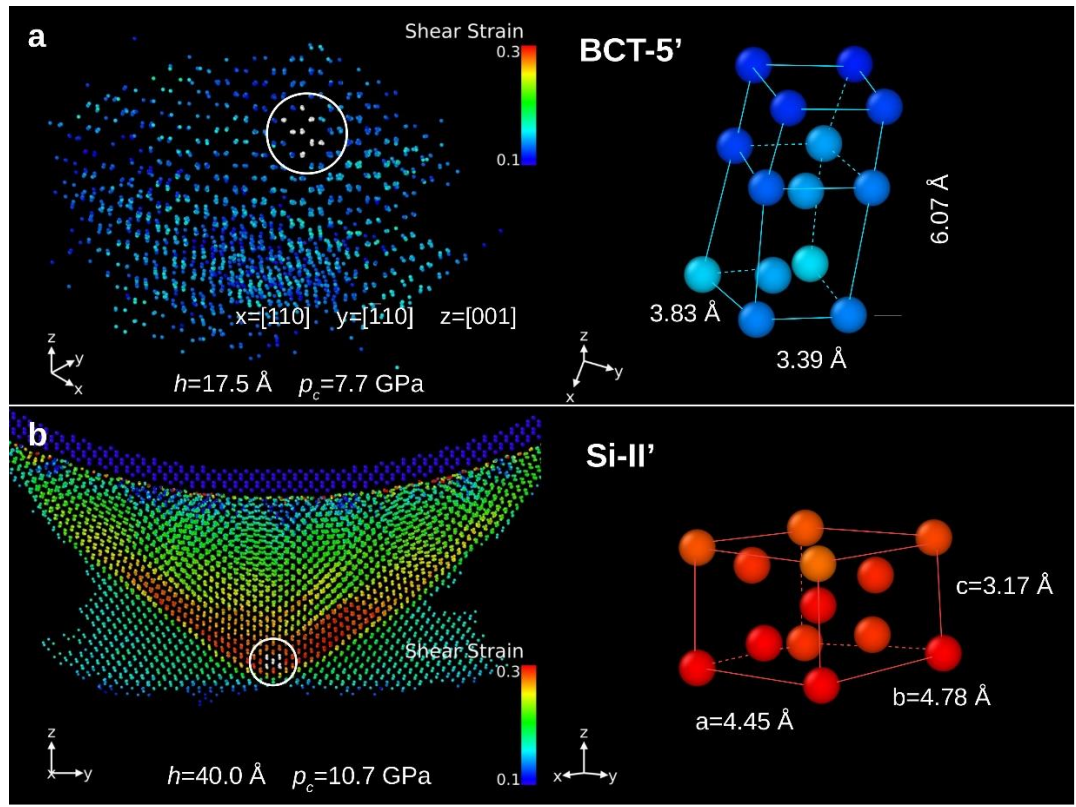

**Fig. 7.** Modelling with the IP potential: The details of the Si-II' and BCT-5' structures. (a) The example of the BCT-5' phase detection and visualization of its unit cell. (b) Location of the Si-II' phase in the deformed crystal. The displayed Si-II' phase is characterized by low value of c/a≈0.66, while the perfect Si-II structure is defined by the c/a ratio of 0.55.

The crystalline Si-I→Si-III' phase transformation predicted by our MD-simulations with the integrated IP potential (refer to Fig. 6a - snapshot 3 and 4) requires a separate elaboration. The nucleation of this specific structure occurs at a late stage of nanoindentation in four distinct zones which our visualisation of the deformed atomistic arrangement reveals (Fig. 8a).

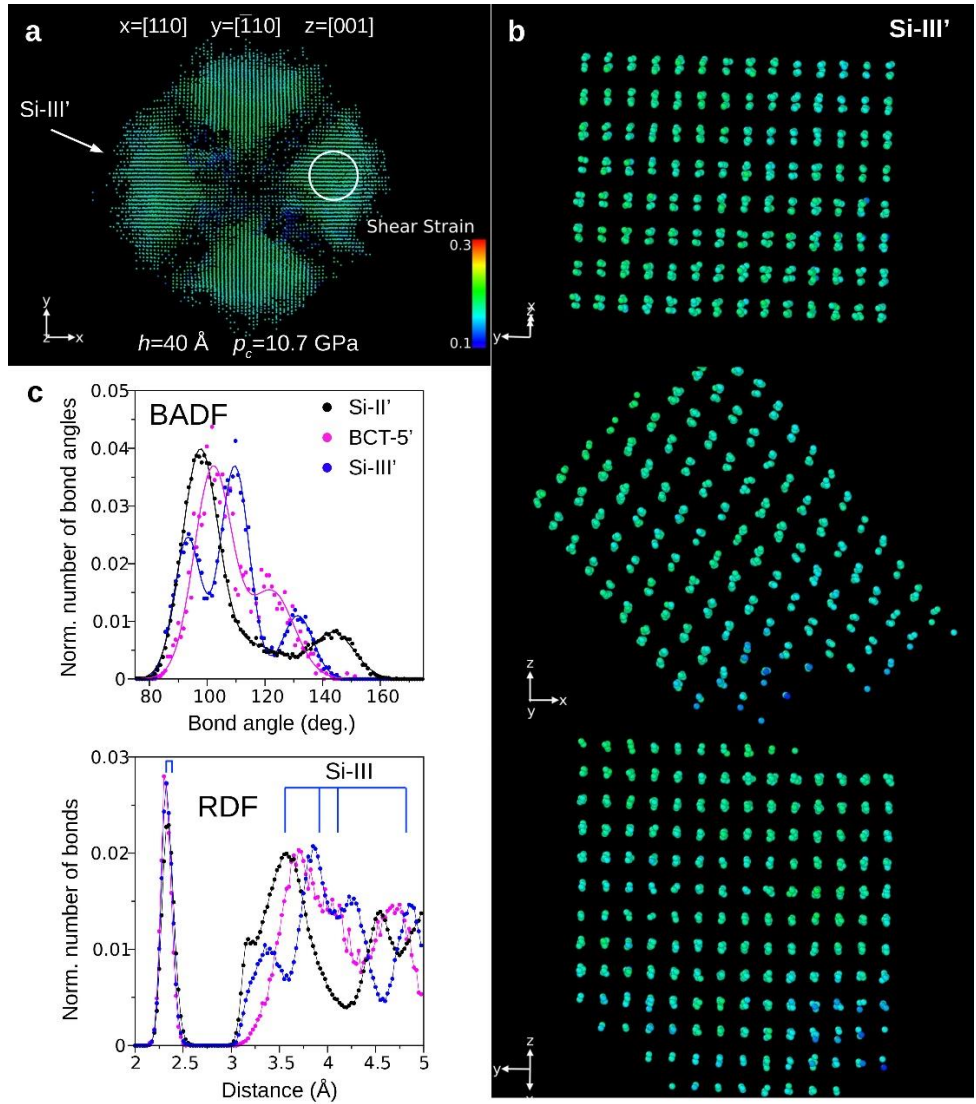

**Fig. 8.** Modelling with the integrated IP potential: The details of Si-III' structure (a) The volume of the Si-III' phase consists of four separate parts. (b) Three orthogonal views of the Si-III' structure selected from the region marked with a white circle. (c) The results of the BADF and RDF analysis show the differences between the three high-pressure Si phases modelled by integrated potential. The RDF of Si-III' phase exhibit peaks cantered at distances near those expected for real Si-III phase.

In order to obtain a clear image of the detected Si-III' structure, we have examined three mutually orthogonal projections of the lattice (Fig. 8b), whose uniqueness is confirmed by the BADF and RDF functions (Fig. 8c). The characteristic BADF peaks of the Si-III' phase are located at 94º, 109º and 132º, while those of RDF are positioned at 3.4, 3.9, 4.3 and 4.9 Å. The RDF data is thus consistent with the theoretical spectrum for the Si-III phase (Fig. 8c), while the additional peak at 132º in the BADF spectrum represents a minor deviation. The overall conclusion is that the phase we are dealing with here is roughly equivalent to the Si-III (see Fig. S5). The unit cell of the Si-III' phase has been presented in Supplementary Materials (Fig. S6), and provides further proof of the IP's viability.

The obvious advantage of the IP potential for MD-simulations lies in the prospect of modelling both phase transformations and dislocation processes simultaneously. Fig. 9 presents nucleation and a well-developed dislocation structure captured at the indentation depth of $h$=43 Å ($p_c$=11.4 GPa) and $h$=45 Å ($p_c$=11.5 GPa) together with high-pressure phases that exist at this stage of nanodeformation of an Si crystal. A DXA analysis of the atomistic arrangement enables us to identify the dislocation lines in the $\{111\}$ planes, and to define their Burgers vectors as $\vec{b} = 1/2\langle 110 \rangle$. Slip planes bounded by dislocation lines are also clearly visible. It is also worth noting that the dislocations assume the correct orientation, namely, downwards into the crystal volume, which agrees with experimental observations [42,48,93]. The most interestingly the dislocations do not develop parallel to the indented surface, contrary to the scenario imposed by the simulations with other potentials, while the calculations with T2 potential did not mirror the lattice defects activity at all.

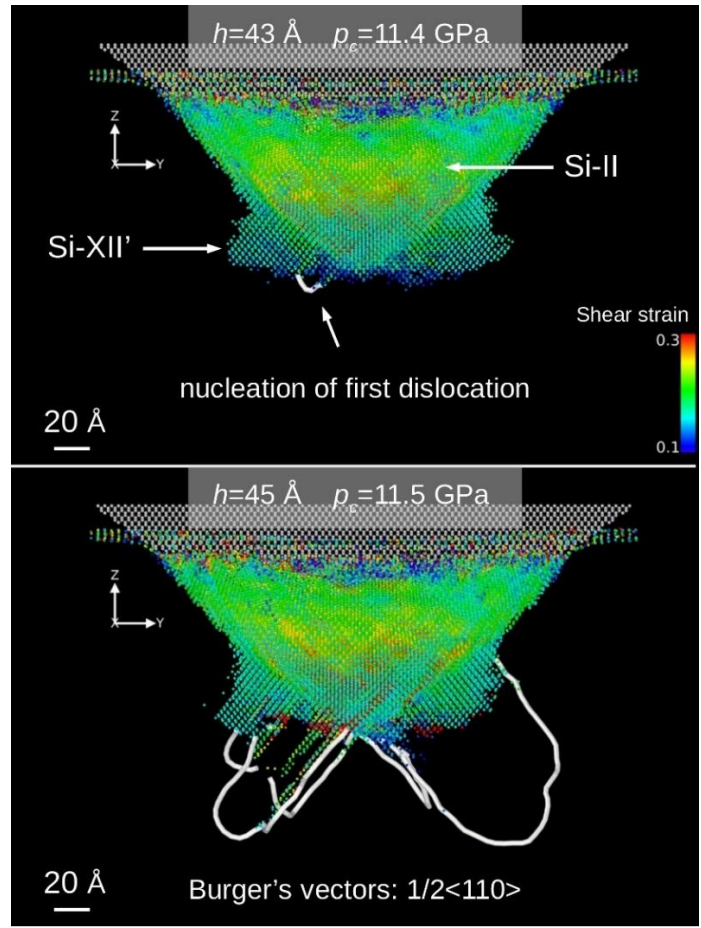

**Fig. 9.** Modelling with the integrated IP potential: The visualization of the 'pattern of displacements' that reveals nucleation and development of dislocations. The dislocation lines are situated in the {111} planes with the Burgers vectors $\vec{b} = 1/2\langle 110 \rangle$.

## 4. Discussion

Our evidence consists of two complimentary parts: the first puts forward a method of constructing a integrated interatomic potential; the second demonstrates that MD simulations of Si nanodeformation with the IP potential are able to mimic the mechanical response of Si more faithfully than other potentials in common use. In general, despite its simplicity the introduced procedure of selecting a universal and robust potential meets all the criteria stipulated by Harrison *et al*. [32] such as simplicity, accuracy, enhanced transferability and computational speed. Significantly, our integrated IP potential inherits the characteristics of its constituent parts: the elastic constants we obtained for the Si-I structure (Table 1) follow the weighted sum of the ones derived by means of the SW and T2 potentials. By comparing the outcome with that of T3s - presently the most advanced model for studying the mechanical behaviour of nano-strained Si crystal [65,71] – we demonstrate that the IP result represents a better match with the experimental reality.

That the integrated SW-T2 potential preserves the characteristics of its constituent parts becomes evident in MD simulations of the structural changes which occur in Si during spherical nanoindentation. The results of our calculations indicate that modelling with the IP allows the capture of the subtle interplay between phase transformation and the generation of structural defects, which the T2 and SW potentials can also model, each in its own right. However, the SW potential, while performing well at modelling dislocation activity, fails to reflect the high-pressure crystalline phases of Si. Similarly, the T2 potential, which is only used to model the Si-I→Si-II' phase transformation (Fig. 2). Whether plastic deformation of Si is initiated by defect generation or phase transformation has been a matter of controversy for decades [17,20,22,40-48,64,70-72,94,95].  For instance, Wong *et al*. [22,44] maintain that the irreversible deformation of an Si crystal is induced by spherical indentation (tip diameter of $\phi$ = ~21.5 μm) is of "a stochastic kind", and thus seems to be caused either by phase transition or the generation of lattice defects. All the same, their work is worthy of attention, since it approaches the issue of incipient plasticity in nano-indented Si, in a methodical and persuasive manner. Accordingly, Wong *et al*. witnessed Si plasticity being steered by the transformation in a compact mass adjacent to the acting tip [22,44], which is also borne out by our simulations with the IP potential (Fig. 6 and Fig. 7).

To be more precise, our MD-prediction envisages the formation of a single, relatively large and dense volume of what we term the Si-II' phase (see the lattice in Fig. 7b) directly under the acting spherical indenter. This is in contrast to the results obtained using T3s potential, which generates isolated areas of the Si-II structure ($\beta$-tin configuration – Fig. 4a and Fig. S3) distributed within narrow slip bands. The generation of the Si-II' structure modelled by means of the integrated IP potential (Fig. 6) is volumetric in character as opposed to local, such as the formation of the Si-II dispersed within V-shaped slip bands (Fig. 4). It is therefore reasonable to assume that the structural transformation of the initial Si-I (diamond, cubic) into the high-pressure Si-II (Fig. 4) is governed predominantly by sheer stress, whereas that from the Si-I into Si-II' (Fig. 7) by hydrostatic one.

*In situ* observations of the structural changes in a nanoindentation-deformed Si crystal have been performed by means of the Raman spectroscopy [40-42] and combined with a microscopic inspection and structural analysis of the material close to the residual indents [17,20,22,43-46]. Thus, Gerbig *et al*. [41] have been able to provide experimental proof that the Si-I→Si-II transition detected during the loading cycle does not proceed in a direct but gradual manner, since it is accompanied by the formation of the *dc-2* structure, which has been referred to as a 'severely distorted Si-I phase'. The *dc-2* had earlier been mistaken [40] for the BCT-5 based on MD-simulations with the non-screened Tersoff type potential [88,89].

The results of our calculations with the hybrid HP1 and T3s potentials point to the formation of the Si-II' and Si-II, respectively. In each case, this is preceded by the initial transition from the Si-I to a high pressure BCT-5' one. The ambiguity contained in the description of the simulated BCT-5' phase (Fig. 3b and Fig. 6) parallels that of the Raman spectra found by Gerbig *et al*. [40,41]. It is worth emphasising, however, that the BCT-5 phase does not show up for late stages of indentation in simulations performed with IP potential (Fig. 6) as opposed to the T3s (Fig. 4). The result is supported by a recent Raman spectroscopy examination of nanoindentation-deformed Si crystals [41,42].

Our main point of reference with regard to the IP is the Tersoff-type screened T3s potential, which to this day represents the most advanced approach to the complex nanodeformation of Si crystals, accounting both for dislocation activity as well as stress-induced phase transformations [70-72]. Thus, both the T3s- (Fig. 5) and IP-driven MD-simulations (Fig. 9) reveal dislocation activity that contributes to the overall indentation deformation in addition to phase transformations (Fig. 3, Fig. 4 and Fig. 6). However, there are some crucial differences between the two. In the scenario obtained with the integrated IP potential, slip leads to the generation of large, extended linear defects on the {111} planes that develop outwardly of the Si volume occupied by high pressure Si-II'/Si-XII' phases (Fig. 9). The situation looks different in the case of the T3s potential where slip, while also occurring along the {111} planes, generates limited dislocations in the direction parallel to the (001) surface of the deformed Si crystal (compare Fig. 5 and Fig. 9). This remarkable difference is related to the energy consumption in a nano-indentation deformed Si crystal during its irreversible deformation.

The total elastic energy $E_{el}$ stored in a deformed crystal consists of a volumetric part $E_{el,h}$ - associated with the hydrostatic stress, and $E_{el,s}$ - related to the acting shear stress. The volumetric character of the Si-I→Si-II' transformation deduced with the IP potential suggests that the process is mainly at the expense of $E_{el,h}$'s. Consequently, the remaining $E_{el,s}$ part is used up for the generation of dislocations. By contrast, the shear stress-steered Si-I→Si-II transformation, modelled with the T3s potential consumes the lion's share of the $E_{el,s}$ energy, resulting in its deficiency for the generation of a sizable dislocation network (Fig. 5) that could be observed experimentally (*e.g.*, Figs. 2-3 in Ref. 48 and Fig. 4 in Ref. 93).

Another point worth noting is that the attributes of the hybrid IP potential enabled the detection of the high-pressure Si-III' structure already in the late stages of the loading cycle (Fig. 6 - snapshot 4 and Fig. 7). Remarkably, it is similar to the Si-III (bc-8) arrangement detected experimentally in the course of unloading, however [40,94]. We have been able to come across only one publication to locate the Si-III structure in the loading stage, namely by Gerbig *et al*. [42], where it appears in the Raman spectrum taken at the load P=50 mN (see Figs. 2 and 3 in Ref. 42).

## 5. Conclusions

In summary, we have managed to develop a reliable integrated interatomic IP potential with remarkable readiness which enables modelling of the phenomena that occur in nanoindentation-deformed Si crystals in close proximity to earlier experiments. MD simulations with IP provide a realistic contact stress at the onset of a plastic deformation of silicon and correctly predict the essential role of a high-pressure BCT-5' structure formation in all its crystallographic complexity (as a highly-compressed Si-I structure or a distorted BCT-5 phase). Moreover, an integrated potential makes it possible - again in accordance with earlier experiments - to reflect the subsequent stages of Si nanodeformation such as a further transformation to the Si-II' phase (c/a≈0.66), the formation of an Si-III-like structure as deformation proceeds and, finally, the activation of dislocation processes. Furthermore, our independent verification of the integrated potential IP-2 constructed for germanium, offered a successful solution to MD simulations of liquid-to-amorphous Ge phase transformation (details in Appendix). The potentials integration procedure appears more accessible than the more traditional methods, and, last not least, the applied procedure thus asserts itself as a viable answer for a range of materials (Supplementary Note 2).

## Appendix

### MD-modelling of germanium solidification using the integrated Stillingr-Weber/Tersoff potential

In order to examine wider relevance of our approach, our potential-integration method was subsequently applied to germanium. The intriguing question was whether it could mirror the structural changes in germanium during the solidification process [96]. Needless to say, the liquid-to-amorphous Ge transformation and changes in crystalline materials such as silicon that we had dealt with earlier are two kinds of effect of an entirely different order. Similarly to silicon, we began with an examination of variously weighted versions of integrated potential, arriving at a combination: $E^{T2-SW} = 0.2E^{T2} + 0.8E^{SW}$, expected to model simultaneously the liquid and amorphous structures of Ge (Supplementary Table S1 and Table S2). As a next step, we examined the IP-2 potential's aptitude for reproducing the liquid-to-amorphous transformation observed for Ge during rapid cooling [97]. The profile of the radial distribution function (RDF) obtained with IP-2 potential matches the experimental data for both the liquid [97, 98] and amorphous [99,100] Ge phase (Fig. App.1). At the same time, neither SW nor Tersoff potential on its own can faithfully recapture the structural changes of a quickly solidifying germanium fully apprehends the structural changes of a rapidly solidifying germanium.

The Stillinger-Weber potential reflects the liquid Ge phase (Fig. App.1A), yet fails to reproduce the amorphous one, which is revealed by the erroneous location ($r \cong 3.0$ and 3.5 Å) of the characteristic RDF-peaks (Fig. App.1B). In contrast, applying the Tersoff potential results in a properly modelled amorphous phase (Fig. App.1B) and a poorly defined liquid one, as shown by the unexpectedly low RDF minimum at $r \cong 3$ Å (Fig. App.1A). The same unrealistic peaks do not however show up in the RDF profiles determined with the hybrid HP2 potential (Fig. App.1). The results demonstrate the hybrid potential's ability to reflect both the liquid and amorphous phase of Ge with equal precision, proving the IP-2 is well suited to describing germanium's liquid-to-amorphous transition.

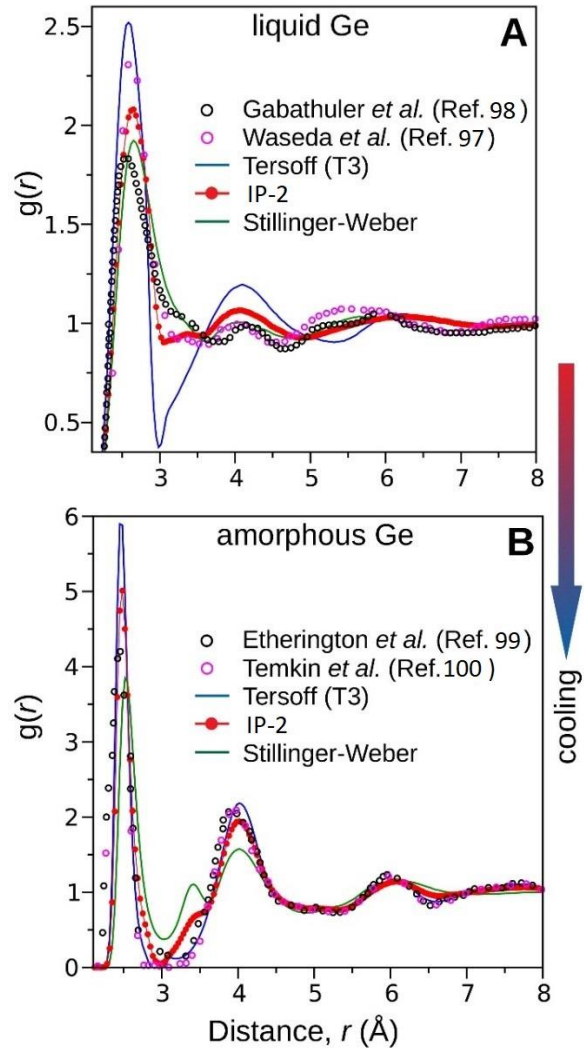

**Fig. App. 1**  MD simulations of the structural changes in Ge during solidification using a hybrid in contrast to other potentials. (**A**) RDF spectra obtained for liquid Ge with the Stillinger-Weber (green), Tersoff (violet) and the hybrid HP2 (red) potentials set against experimental data by Waseda *et al.* [97] (magenta dotted line) and Gabathuler *et al.* [98] (black dotted line). The results favour the Stillinger-Weber and HP2 potential s as suitable to simulate the Ge liquid phase (the peaks obtained with Tersoff formula appear unrealistic. (**B**) RDF spectra obtained for solid Ge with Stillinger-Weber (green), Tersoff (violet) and the HP2 (red) potentials set against experimental results by Etherington *et al.* [99] (black dotted line) and Temkin *et al.* [100] (magenta dotted line).

**Data availability**

All relevant data supporting the key findings of this study are available within the article and its Supplementary Information files or from the corresponding author upon reasonable request.

**Declaration of Competing Interest**

The authors declare that they have no known competing financial interests or personal relationships that could have appeared to influence the work reported in this paper.

**Acknowledgments**

RA thanks Lifang Zhu and Jan Janssen for access to their programs essential to the calculation of the melting point of Ge with diamond structure. RN is grateful to Toshihiro Shimada (Hokkaido University) and Adam Poludniak (Seirei University) for important discussions and longstanding support. This research was assisted by the Academy of Finland - Research Platform OMA for Programmable Materials (The Consortium PROPER). All computer simulations used resources provided to the Nordic Hysitron Laboratory by the CSC-IT Centre for Science, Finland, which we gratefully acknowledge. DC is grateful for the support from the National Science Centre, Poland (Grant No. 2016/21/B/ST8/02737). RN appreciates the visiting scholar opportunity at Hokkaido University.

**Supplementary materials**

Supplementary material associated with this article can be found, in the online version, at … doi???

**Supplemental for**

# Comprehensive structural changes in nanoscale-deformed silicon modelled with an integrated atomic potential

Rafal Abram[a], Dariusz Chrobak[b],Jesper Byggmästar[c], Kai Nordlund[c], Roman Nowak[a,d,*]

[a]  Nordic Hysitron Laboratory, School of Chemical Engineering, Aalto University, Espoo, 00076 Aalto, Finland
[b]  Institute of Materials Science, University of Silesia, 41-500 Chorzow,75 Pulku Piechoty 1A, Poland
[c]  Department of Physics, University of Helsinki, P.O. Box 43, FI-00014, Finland
[d]  Division of Materials Chemistry, Faculty of Engineering, Hokkaido University, Sapporo 060-8628, Japan

*Corresponding author: Roman Nowak   Email: roman.nowak@aalto.fi

**Contents:**

Supplementary Text

**Supplementary Note 1: The integrated T2-SW (IP) potentials and Lammps**

In order to implement any hybrid potential in the *Lammps* simulation package, the parameters of the SW and T2 component potentials are required that are modified according to equations:

$$E^{\text{T2-SW}} = w_{\text{T2}} E^{\text{T2}} + w_{\text{SW}} E^{\text{SW}}$$

$$w_{\text{SW}} + w_{\text{T2}} = 1$$

where: $w_{\text{SW}} = 0.28125$ and $w_{\text{T2}} = 0.71875$ define our integrated IP model for silicon

The files that define pertinent potential parameters are as follows:

*The case of SW component of the IP potential*

# file: Si.sw.28125
# element 1, element 2, element 3, epsilon, sigma, a, lambda, gamma, costheta0, A, B, p, q, tol
# Stillinger and Weber,  Phys. Rev. B, v. 31, p . 5262, (1985)
# lambda: 5.90625 = 21.0 * 0.28125
# A: 1.982687703 = 7.049556277 * 0.28125
Si Si Si 2.1683  2.0951  1.80  5.90625  1.20  -0.333333333333  1.982687703  0.6022245584  4.0  0.0 0.0

*The case of T2 component of the IP potential*

# file: Si.tersoff.71875
# element 1, element 2, element 3,  m, gamma, lambda3, c, d, costheta0, n, beta, lambda2, B, R, D, lambda1, A
# Tersoff, Phys Rev B, 37, 6991 (1988)
# A: 2346.503125 = 3264.7 * 0.71875
# B: 68.54934375 = 95.373 * 0.71875
Si  Si   Si  3.0 1.0 1.3258 4.8381 2.0417 0.0000 22.956 0.33675  1.3258  68.54934375  3.0  0.2  3.2394 2346.503125

**Supplementary Note 2: The nanoindentation $p_c$-$h$ loading curves obtained with various interatomic potentials**

A comparison of the relationship between contact pressure and indentation depth ($p_c$-$h$) determined by MD-simulations using the IP, Tersoff T2 and T3s potentials (Fig. S1) reveals the advantage of our integrated approach. Thus, the contact pressure of 12.6 GPa at the onset of an irreversible deformation obtained with the T3s is grossly overestimated when compared with the value of just 6.2 GPa yielded by nanoindentation tests combined with Raman spectroscopy [S1]. Far more realistic appears to be the level of approx. 7.7 GPa obtained with IP potential.

**Supplementary Note 3: Effectiveness of MD simulations with integrated IP potential**

In order to illustrate the advantage of the hybrid approach, we have estimated the effectiveness of the 'GAP' potential for silicon produced by employing the 'machine-learning procedure' [S6], using the supercell (10x10x10 nm) of silicon with a diamond structure (8 nodes x 128 mpi processes, 8000 atoms, 100 time steps) requiring 38.9 s of calculation time. The cluster used in our simulations of Si nanoindentation with a hybrid potential contained 1200000 atoms. To secure system equilibrium, each indenter shift-increment required at least 5000 stages (time-increment). This amounts to approx. 82 hours needed to accomplish a single shift of the tip. With at least 100 displacements in our simulations with our HP1 potential, the GPA potential would require a tremendous amount of machine time.

**Supplementary figures (Figs. S1 ~ S6)**

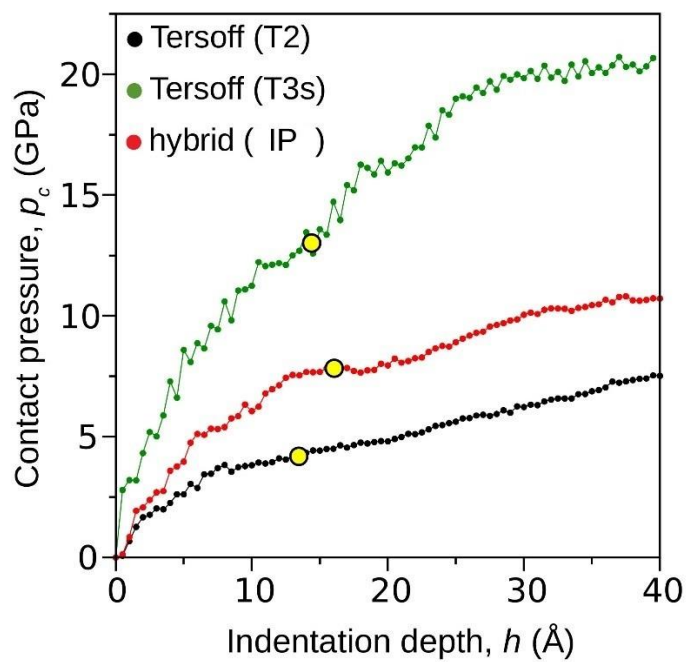

**Fig. S1. The contact stress versus depth ($p_c$-$h$) indentation loading curves derived from MD simulations with the T2, T3s and HP1 potentials.** The onset of the stress-induced phase transformation is marked in yellow. Compared with the T3s potential, the IP potential does not result in an overestimation of the contact stress.

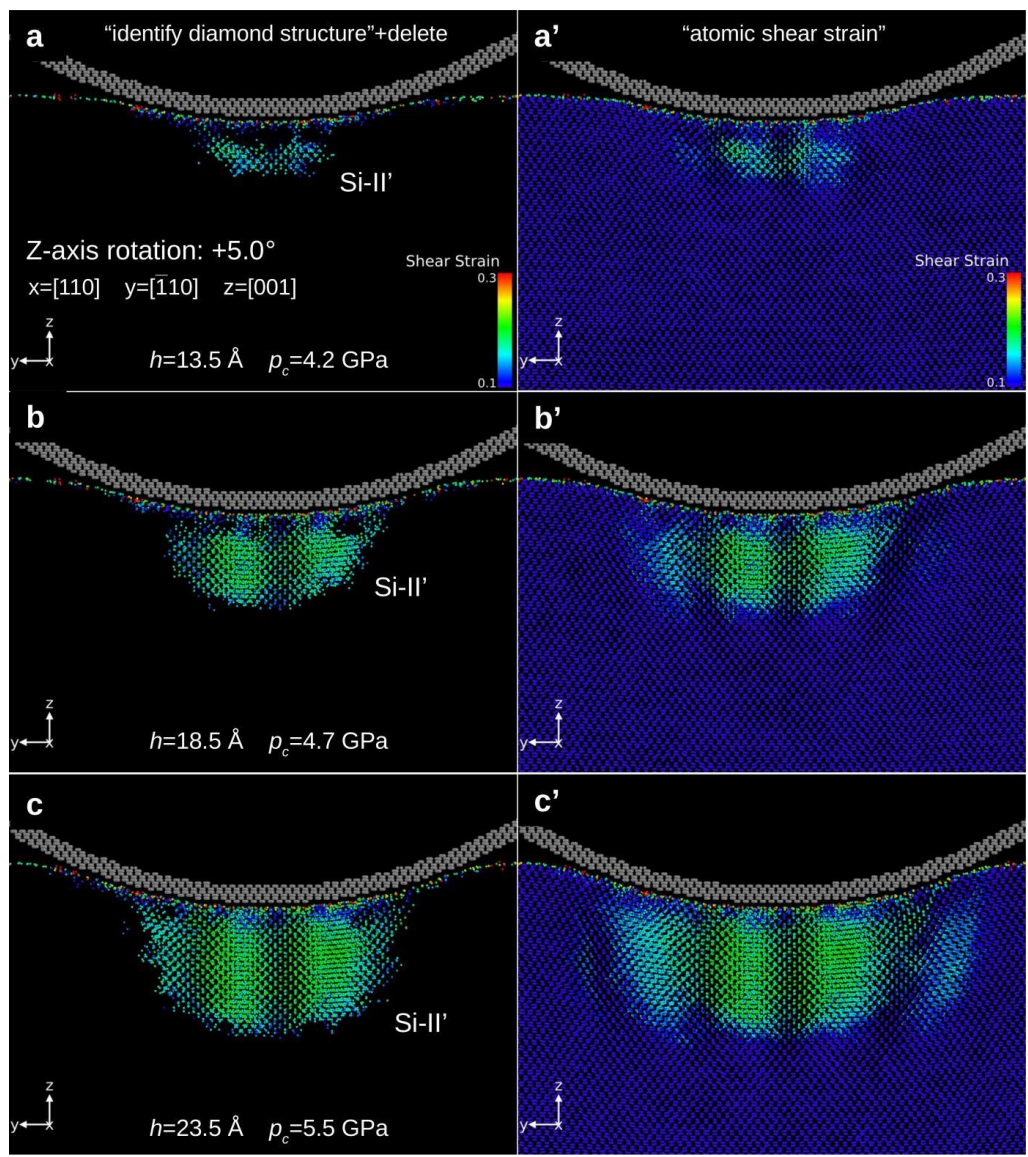

**Fig. S2** Modelling with the T2 potential: Evolution of the high pressure Si-II' phase.
(a), (b), (c) Selected indentation stages ($h$=13.5, 18.5 and 23.5 Å) showing nucleation and further development of the Si-II' phase. The visualization obtained by means of the "*identify diamond structure*" modifier of the OVITO software. (a'), (b'), (c') The structural changes in Si obtained by means of the "*atomic shear strain*" modifier of the OVITO software. The volume occupied by the high-pressure Si-II' phase coincides with the area of the enhanced level of shear strain

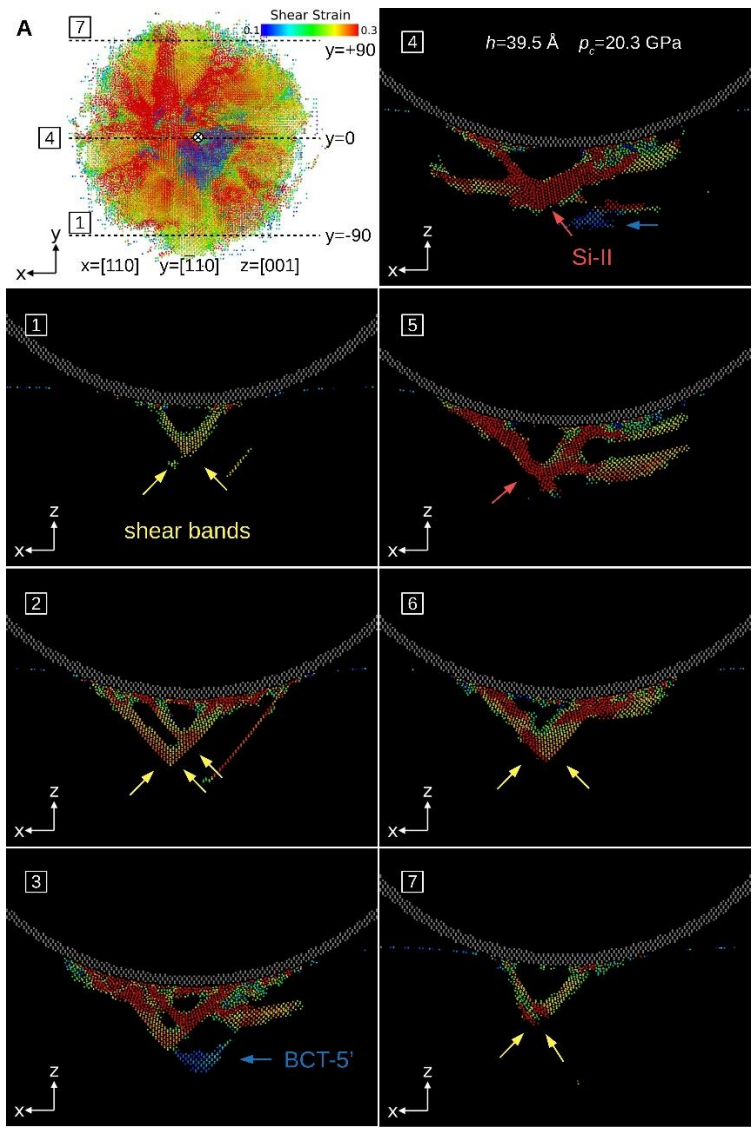

**Fig. S3. The distribution of the high-pressure Si phases modelled with the T3s potential.**
(A) Cross-sections (5 Å thick lamellae defined by seven y-axis coordinates) of the high-pressure Si phases formed under a rigid spherical indenter ($h$=39 Å, $p_c$= 20.3 GPa). A set of snapshots 1~7 (y = -60, -40, -20, 0, +20, +40 and +60) demonstrate that high-pressure phases do not represent a compact volume adjacent to the contact surface, in contrast to the prediction with the IP potential (*e.g.*, Fig. 2 and Fig. S2) or earlier experimental observations, *e.g.* [S2, S3, S4]. The effects associated with shear-bands are marked in yellow, while the formation of BCT-5' and Si-II structure in blue and red, respectively.

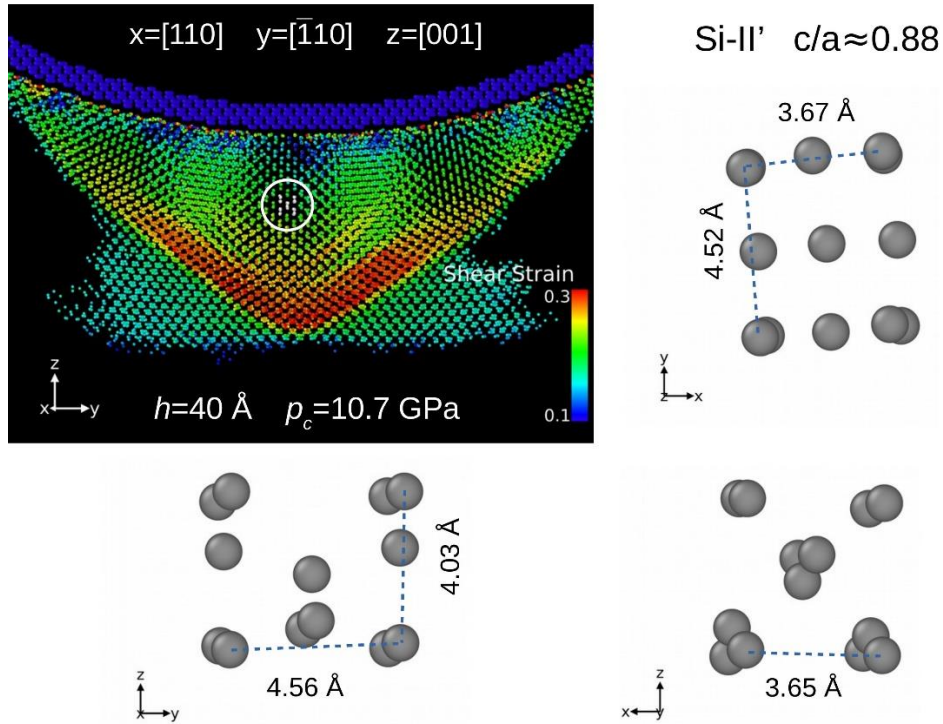

**Fig. S4. Modelling with the IP potential: The characteristics of the Si-II' structure.** A variously oriented Si-II' phase unit cell indicates a higher value of the tetragonal parameter $c/a$ in places where shear stresses/strains are smaller. The lowest level of the $c/a$ parameter was found in shear stress concentrations (marked in red).

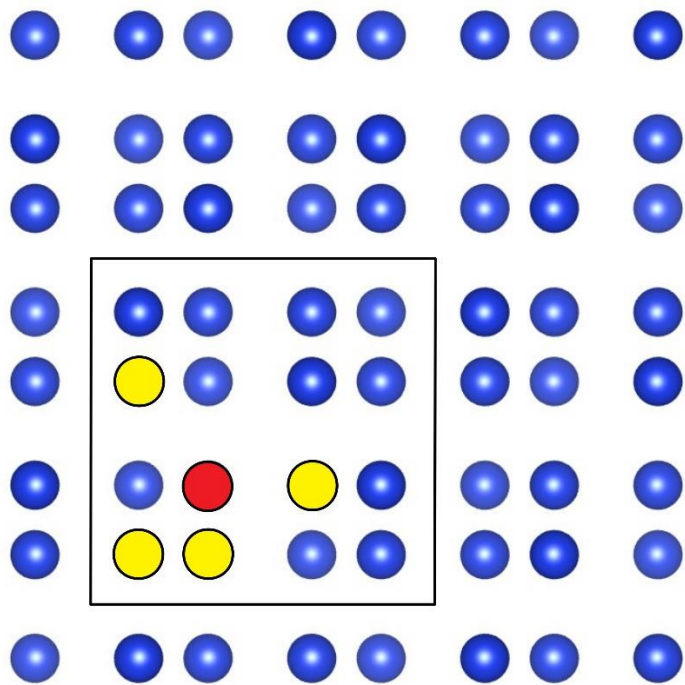

**Fig. S5. The cubic unit cell of the Si-III phase**. The structure named otherwise bc-8 after Crain *et al.* [S5] is visualized using the VESTA software (jp-minerals.org/vesta/en/). The Si-III phase is characterized by the lattice parameter of *a*=6.64 Å. Each of Si atoms (*e.g.*, red) has four nearest neighbours (yellow). The distance to the neighbours equals 2.39 or 2.31 Å, while the bond angles are 99º and 118º.

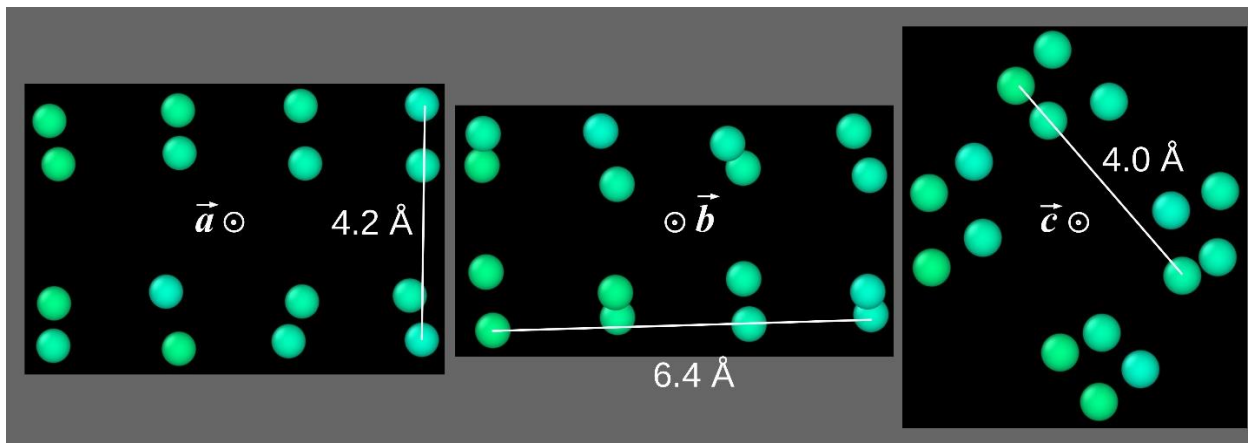

**Fig S6. A set of orthogonal projections of the Si-III' unit cell obtained with the IP potential.** The unit vectors $\vec{a}$=[0.63,-0.01,-0.78], $\vec{b}$=[0.56,0,0.83] and $\vec{c}$=[0.17,-0.98,-0.1] are perpendicular to the figure's plane and point outwards. The indentation axis is defined along the vector $\vec{w}$=[0,0,1]. The angles between the unit vectors read: ∡AB=107º, ∡AC=91º and ∡BC=100º.

# Supplementary tables (Table S1 ~ S2)

**Table S1** The elastic parameters $X_i$ calculated for selected Ge interatomic potentials, namely, the Stillinger-Weber (SW), Tersoff (T2) and integrated IP-2. The lattice constant for the modelled *diamond cubic* phase of Ge equals 5.657 Å.

| $X_i$ [GPa] | SW | T2 | $w_{SW}X_{SW}+w_{T2}X_{T2}$ | IP-2 | Experimental data** |
|---|---|---|---|---|---|
| $B$ | 80,0 | 75,8 | 79.2 | 79.2 | 75 |
| $c_{11}$ | 117,8 | 138,5 | 121.9 | 121.9 | 126.0 |
| $c_{12}$ | 61,2 | 44,4 | 57.8 | 57.8 | 44.0 |
| $c_{44}$ | 41,1 | 69,5 | 46.8 | 42.8 | 67.0 |

** http://www.ioffe.ru/SVA/NSM/Semicond/Ge/mechanic.html

**Table S2** The melting points calculated for selected Ge interatomic potentials: Stillinger-Weber (SW), Tersoff (T2) and integrated IP-2. Also included is the weighted sum of the melting points: $w_{SW}X_{SW}+w_{T2}X_{T2}$.

| SW | T2 | $w_{SW}X_{SW}+w_{T2}X_{T2}$ | IP-2 | commonly accepted value* |
|---|---|---|---|---|
| 1295 | 2500 | 1536 | 1679 | 1210 |

* http://www.ioffe.ru/SVA/NSM/Semicond/Ge/thermal.html

**Supplementary References**